\documentclass[twocolumn,showpacs,aps,prl,superscriptaddress]{revtex4}

\usepackage{graphicx}
\usepackage{dcolumn}
\usepackage{amsmath}
\usepackage{epsfig}

\input pubboard/babarsym

\newcommand{\BABARPubNumber}  {03/032}
\newcommand{\SLACPubNumber} {10220}

\def\figurebox#1#2#3{%
    \def\arg{#3}%
    \ifx\arg\empty
    {\hfill\vbox{\hsize#2\hrule\hbox to #2{\vrule\hfill\vbox to #1{\hsize#2\vfill}\vrule}\hrule}\hfill}%
    \else
    {\hfill\epsfbox{#3}\hfill}%
    \fi}

\def\btodk   {\ensuremath {B^-{\to}D^0 K^-}}

\def\btodkcpm {\ensuremath {B^-{\to}D^0_{\CP} K^-}}
\def\btodkcpp {\ensuremath {B^+{\to}D^0_{\CP} K^+}}
\def\btodkcp {\ensuremath {B^\pm{\to}D^0_{\CP} K^\pm}}
\def\btodp   {\ensuremath {B^-{\ra}D^0 \pi^-}}

\def\btodpcpm {\ensuremath {B^-{\to}D^0_{\CP} \pi^-}}

\def\btodh   {\ensuremath {B^-{\to}D^0 h^-}}

\def\btohpp   {\ensuremath {B^-{\to}h^- \pi^-\pi^+}}
\def\btokkk   {\ensuremath {B^-{\to}K^- K^-K^+}}

\def\dotokp  {\ensuremath {D^0{\ra}K^-\pi^+}}
\def\dotokppp  {\ensuremath {D^0{\ra}K^-\pi^+\pi^+\pi^-}}
\def\dotokppo  {\ensuremath {D^0{\ra}K^-\pi^+\pi^0}}
\def\dotopp  {\ensuremath {D^0{\ra}\pi^-\pi^+}}
\def\dotokk  {\ensuremath {D^0{\ra}K^-K^+}}

\begin{document}


\begin{flushleft}
\babar-PUB-\BABARPubNumber\\
SLAC-PUB-\SLACPubNumber\\
\end{flushleft}

\title{
{\large \bf
Measurement of the branching fractions and \CP-asymmetry of
$B^-\ra D^0_{(\CP)}K^-$ decays with the \babar\ detector}
}

%
\author{B.~Aubert}
\author{R.~Barate}
\author{D.~Boutigny}
\author{J.-M.~Gaillard}
\author{A.~Hicheur}
\author{Y.~Karyotakis}
\author{J.~P.~Lees}
\author{P.~Robbe}
\author{V.~Tisserand}
\author{A.~Zghiche}
\affiliation{Laboratoire de Physique des Particules, F-74941 Annecy-le-Vieux, France }
% ...
\collaboration{The \babar\ Collaboration}
\noaffiliation

\date{\today}

\begin{abstract}
We present a study of \btodkcpm decays, where $D^0_{\CP}$ is
reconstructed in \CP-even channels,
based on a sample of 88.8 million $\Upsilon(4S)\rightarrow B\bar{B}$ decays
collected with the \babar\ detector at the PEP-II \epem storage ring.
We measure the ratio of Cabibbo-suppressed to Cabibbo-favored branching
fractions
$\BR(\btodkcpm)/\BR(\btodpcpm)= (8.8\pm 1.6\stat\pm 0.5\syst)\times 10^{-2}$
and the \CP asymmetry $A_{\CP}= 0.07\pm 0.17\stat\pm 0.06\syst$. We also
measure $\BR(\btodk)/\BR(\btodp)=(8.31\pm0.35\stat\pm0.20\syst)\times
10^{-2}$ using a sample of 61.0 million \BB pairs.
\end{abstract}


\maketitle

The recent observation of \CP\ violation in the $B$ meson
system~\cite{sin2b_babar} has provided a clean measurement of the angle
$\beta$ of the unitarity triangle. Although this measurement is in good
agreement with the expectations of the Standard Model derived from other
measurements of weak interactions, further
measurements of \CP\ violation in $B$ decays are needed to
overconstrain the unitarity triangle and confirm the CKM mechanism or
observe deviations from it. A theoretically clean measurement of the
angle $\gamma$ can  
be obtained from the study of $B^-{\to}D^{(*)0}K^{(*)-}$ decays by
reconstructing the $D^0$ meson in Cabibbo-allowed \CP eigenstates and 
doubly Cabibbo-suppressed decays~\cite{gronau1991}. 

In this Letter we present the measurement of the ratios of
Cabibbo-suppressed to Cabibbo-favored branching fractions 
\begin{equation}
R_{(\CP)}=\frac{\BR(B^-{\ra}D^0{(\Dz_{\CP})}K^-)+\BR(B^+{\ra}\bar{D^0}(\Dz_{\CP})K^+)}{\BR(B^-{\ra}D^0(\Dz_{\CP})\pi^-)+\BR(B^+{\ra}\bar{D^0}(\Dz_{\CP})\pi^+)}
\end{equation}
with $D^0$ reconstructed in Cabibbo-allowed or \CP-even ($D^0_{\CP}$) channels.
The direct \CP asymmetry
\begin{equation}
A_{\CP}=\frac{\BR(\btodkcpm)-\BR(\btodkcpp)}{\BR(\btodkcpm)+\BR(\btodkcpp)}
\end{equation}
is also measured.
The measurement of $R$ uses a sample of 61.0 million
\FourS\ decays  
in $B\overline{B}$ pairs collected with the 
\babar\ detector at the \pep2\ asymmetric-energy $B$ factory. The analysis of 
\btodkcp\ decays uses a sample of 88.8 million $B\overline{B}$ pairs.
Since the \babar\ detector is described in detail
elsewhere~\cite{detector},  
only the components that are crucial to this analysis are
summarized here. 
Charged-particle tracking is provided by a five-layer silicon
vertex tracker (SVT) and a 40-layer drift chamber (DCH). 
For charged-particle identification, ionization energy loss in
the DCH and SVT, and Cherenkov radiation detected in a ring-imaging
device (DIRC) are used. 
Photons are identified 
by the electromagnetic calorimeter
(EMC), which comprises 6580 thallium-doped CsI crystals. 
These systems are mounted inside a 1.5-T solenoidal
superconducting magnet. 
We use the GEANT~\cite{geant} software to simulate interactions of particles
traversing the detector, taking into account the varying
accelerator and detector conditions. 

We reconstruct \btodh\ decays, where the prompt track $h^-$ is a kaon
or a pion (reference to the charge-conjugate state is implied here and
throughout the text unless otherwise stated). Candidates for $D^0$ are
reconstructed in the non-\CP flavor eigenstates $K^-\pi^+$,
$K^-\pi^+\pi^+\pi^-$, $K^-\pi^+\pi^0$ (non-\CP\ modes) and in the
\CP-even eigenstates $\pi^-\pi^+$ and $K^-K^+$ (\CP modes).

To reduce the combinatorial background, only charged
tracks with momenta greater than 150 MeV/c are used to
reconstruct \dotokppp\ and \dotokppo; the prompt particle $h$ is
required to have momentum greater than 1.4 \gevc.
Particle ID information from the drift chamber and, when
available, from the DIRC must be consistent with the kaon
hypothesis for the $K$ meson candidate in all \Dz\ modes and with the pion 
hypothesis for the $\pi^\pm$ meson candidates in the $D^0{\ra}\pi^-\pi^+$
mode. 
For the prompt track to be identified as a pion or a kaon,
we require that its Cherenkov angle be reconstructed 
with at least five photons. We reject a candidate track
if its Cherenkov angle is consistent with that of a proton
or if it is identified as an electron by the DCH and the EMC.

Photon candidates are required to have energies grater than 70 \mev. 
Photon pairs with invariant mass within the range 124--144 \mevcc\ and 
total energy greater than 200 \mev\ are considered \piz candidates.
To improve the momentum resolution, the $\piz$ candidates are kinematically
fit with their mass constrained to the nominal \piz\ mass~\cite{PDG2002}.

The invariant mass of a \Dz\ candidate, $M(D^0)$, must be
within 3$\sigma$ of the mean fitted mass for the channels $K^-\pi^+$,
$K^-\pi^+\pi^+\pi^-$ and $K^-K^+$, and
within 2$\sigma$ for the $K^-\pi^+\pi^0$ channel.
Candidates for \dotopp\ are selected in the range
$1.80<M(D^0)<1.93\ \gevcc$ and the invariant mass of the
$(h^-\pi^+)$ system, where $\pi^+$ is the pion from $D^0$ and $h^-$ is
the prompt track taken with the kaon mass hypothesis, must be greater than
$1.9\ \gevcc$ to reject background from $B^-{\ra}D^0[\ra
  K^-\pi^+]\pi^-$ and $B^-{\ra}K^*[\ra K^-\pi^+]\pi^-$ decays. For all
the \Dz\ decay channels except the $\pi^-\pi^+$ mode a kinematical
fit to the nominal \Dz\ mass~\cite{PDG2002} is applied. 
The \dotopp\ selection differs because of its particular background,
as described later. 

We reconstruct $B$ meson candidates by combining a \Dz\ candidate
with a track $h$. For the non-\CP\ modes, the charge of the
track $h$ must match that of the kaon from the $D^0$ meson decay.
We select $B$ meson candidates by using the beam-energy-substituted mass 
$\mes = \sqrt{(E_i^{*2}/2 + \mathbf{p}_i\cdot\mathbf{p}_B)^2/E_i^2-p_B^2}$
and the energy difference $\Delta E=E^*_B-E_i^*/2$, 
where the subscripts $i$ and $B$ refer to the initial \epem\ system and the 
$B$ candidate respectively, and the asterisk denotes the CM frame. 
The \mes\ distributions for \btodh\ signals are Gaussian distributions centered at
the $B$ mass with a resolution of $2.6 \mevcc$, which does not depend
on the decay mode or on the nature of the prompt track.  
In contrast, the \DeltaE\ distributions depend on the mass assigned to the
prompt track and on the \Dz\ momentum resolution. 
We evaluate $\Delta E$ with the kaon mass hypothesis 
so that the distributions are centered near zero
for \btodk\ events and shifted by approximately $50 \mev$ for \btodp\ events.
The \DeltaE\ resolution is about $20\mev$ for the \dotopp\ mode, and
typically $17\mev$ for the other \Dz\ decay modes. 
We select $B$ mesons in the range $5.2<\mes<5.3\gevcc$ with the
exception of the \dotopp mode, for which \mes is required to be within
3$\sigma$ of the mean value. All $B$ candidates are selected in
the range $-0.10<\Delta E<0.13\gev$. 
For events with multiple \btodh candidates, 
the best candidate is chosen based on the values of $M(D^0)$ and
\mes; this happens in fewer than 1\% of the selected events for two-body
\Dz\ decays and in $\approx 4\%$ of the events for the other \Dz\ decays.

To reduce backgrounds from continuum production of light quarks, we make 
use of two quantities that exploit the different topologies of $\epem \to
 q\overline{q}\ (q=u,d,s,c)$ and \BB\ events. 
The first quantity is the normalized second Fox-Wolfram
moment~\cite{fox_wol}, $R_2\equiv H_2/H_0$,
where $H_l$ is the $l$--order Fox-Wolfram moment of all the charged
tracks and neutral clusters in the event. Only events with $R_2<0.5$ 
are selected. 
The second quantity is the angle $\theta_T$ between the thrust axes
of the $B$ candidate and of the
remaining charged tracks and neutral clusters, evaluated in the CM. 
We require $|\cos\theta_T|<0.9$ for the \dotokp\ mode, and
$|\cos\theta_T|<0.7$ for the \dotokppp\ and \dotokppo\ modes.
For the \dotokk\ and \dotopp\ modes an additional quantity is used 
to suppress further the continuum background: 
the angle $\theta_{Dh}$ between the direction of one of the
decay products of the $D^0$ and the direction of flight of the $B$,
in the $D^0$ rest frame.
The quantities $\cos\theta_T$ and $\cos\theta_{Dh}$ are uncorrelated
for the signal but \emph{not} for the continuum background. This
correlation is 
exploited to make a more efficient cut in the
$\cos\theta_T-\cos\theta_{Dh}$ plane. 

The total reconstruction efficiencies, based on simulated 
signal events, are 42\%($K^-\pi^+$), 14\%($K^-\pi^+\pi^+\pi^-$), 8\%($K^-\pi^+\pi^0$), 34\%($K^-K^+$) and 36\%($\pi^-\pi^+$).  

The main contributions to the \BB background for the non-\CP\ modes come
from the processes $B{\ra}D^{*}h$ ($h=\pi,K$), $B^-{\ra}D^0\rho^-$ and
mis-reconstructed \btodh.  
For $D^0_{\CP}$ decays, the backgrounds $B^-{\ra}K^-K^+K^-$ and
$B^-{\ra}K^-\pi^+\pi^-$~\cite{btokkk} must also be considered, since they have 
the same \DeltaE\ and \mes\ distribution as the $D^0 K^-$ signal.  
The resonant component of these decays is negligible after the selection requirements for the \CP\ modes.

For each \Dz\ decay mode an extended unbinned maximum-likelihood fit
to the selected data events determines the signal and background yields $n_i$
($i=1$ to $M$, where $M$ is the total number of signal and background 
channels). Two kinds of signal events, \btodp\ and \btodk, are considered,
while the number of background sources depends on the $D^0$ channel.
For non-\CP\ modes we consider four kinds of backgrounds: 
candidates selected either from
continuum or from \BB\ events, in which the prompt track is either
a pion or a kaon. In the case of $D^0{\ra}K^-K^+$ we consider two
kinds of background depending on the nature of the prompt
track. Finally, in the case of $D^0{\ra}\pi^-\pi^+$ we consider four
contributions: the $B^-{\ra}K^-\pi^+\pi^-$ and $B^-\ra\pi^-\pi^+\pi^-$
decays and two kinds of generic background depending on the nature of
the prompt track. 

The input variables to the fit for the non-\CP\ and the \dotokk\ modes are \mes, \DeltaE, and a particle identification 
probability for the prompt track based
on the Cherenkov angle $\theta_C$, the momentum $p$ and
the polar angle $\theta$ of the track.
For the \dotopp\ mode, \mes\ is replaced by $M(D^0)$. This allows us to 
separate 
the \btodk\ from the non-resonant $B^-{\ra}K^-\pi^+\pi^-$ contributions since 
the $(\pi^-\pi^+)$-invariant-mass distribution peaks at the $D^0$ mass for 
signal while it is featureless for background. 
The extended likelihood function $\cal L$ is defined as
\begin{equation}
{\cal L}= \exp\left(-\sum_{i=1}^M n_i\right)\, \prod_{j=1}^N
\left[\sum_{i=1}^M n_i {\cal P}_i\left(\vec{x}_j;
\vec{\alpha}_i\right) \right]\,,
\end{equation}
where $N$ is the total number of observed events. 
The  $M$ functions ${\cal P}_i(\vec{x}_j;\vec{\alpha}_i)$ are the
probability density functions (PDFs) for the variables
$\vec{x}_j$, given the set of parameters $\vec{\alpha}_i$. They are
evaluated as a product $\mathcal{P}_i=\mathcal{P}_i(\DeltaE,x)\times\mathcal{P}_i({\theta_C})$ where $x=m_{\rm ES}$ or $M(D^0)$ depending on the 
$D^0$ channel.

The Gaussian shape of the \mes\ PDF for signal events is determined from a pure
sample of \btodp, \dotokp\ decays selected from on-resonance data. 
The \DeltaE\ distribution for \btodk\ signal events is parameterized
with a Gaussian distribution whose parameters are determined from a pure sample of
\btodp\ events selected after the pion mass is assigned to the prompt
track. The displaced \DeltaE\ distribution for \btodp\ is
parameterized with a sum of two Gaussian distributions. The shape of $M(D^0)$ 
is also described by a Gaussian distribution whose parameters are determined 
from data. 

The parameters for the \DeltaE\ and \mes\ distributions for the continuum
background of the non-\CP\ and $D^0{\ra}K^-K^+$ modes are determined 
from off-resonance data. 
The background shape in
\DeltaE\ is parameterized with a linear function, while that of the \mes
is parameterized with an ARGUS threshold function~\cite{argus} 
$f(\mes)\propto \mes\sqrt{1-y^2}\exp[-\xi(1-y^2)]$, where $y=\mes/m_0$
and $m_0$ is the mean CM energy of the beams.
The correlation between \mes\ and \DeltaE\ for the generic \BB\ background
is taken into account with a two-dimensional PDF determined
from simulated events through a method based on the 
Kernel Estimation~\cite{kernel1} technique. For the \dotokk\ mode
the contribution from the non-resonant \btokkk\ decays is
estimated~\cite{btokkk} and added to the continuum and the generic \BB
background. 

The \DeltaE\ and $M(D^0)$ distributions for the continuum and generic
\BB\ background of the \dotopp\ mode are determined from off-resonance
data and simulated events, respectively. The \DeltaE\ distribution is
described by a linear
function while $M(D^0)$ is parameterized with the sum of a linear
function (combinatorial background) and a Gaussian distribution (real \dotopp).
The $M(D^0)$ PDFs of the non-resonant \btohpp\ decays are described by 
linear functions, while the \DeltaE\ distributions are parameterized
with one or two Gaussian distributions, as for the \btodh\ signals.

Finally, the parameterization of the particle identification PDF is performed 
by fitting with a Gaussian distribution the background-subtracted
distribution of the difference between the reconstructed and expected
Cherenkov angles of the kaons and pions from \Dz\ decays, in a pure
$D^{*+}\to\Dz\pip$ ($\Dz{\to}\Km\pip$) control sample.

\begin{figure}[!htb]
\begin{center}
\includegraphics[width=7.5cm,height=3.4cm]{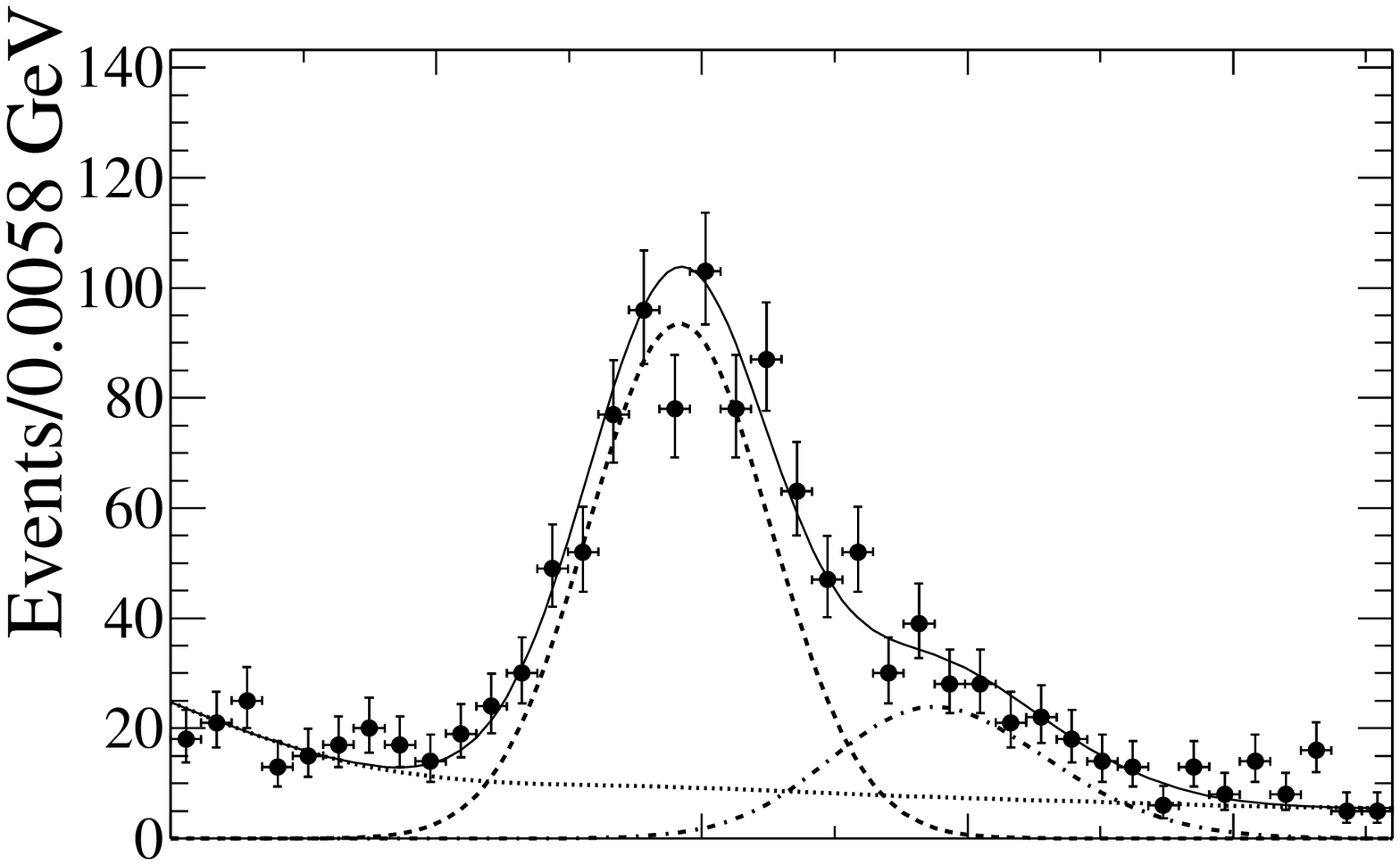}
\includegraphics[width=7.5cm,height=4.0cm]{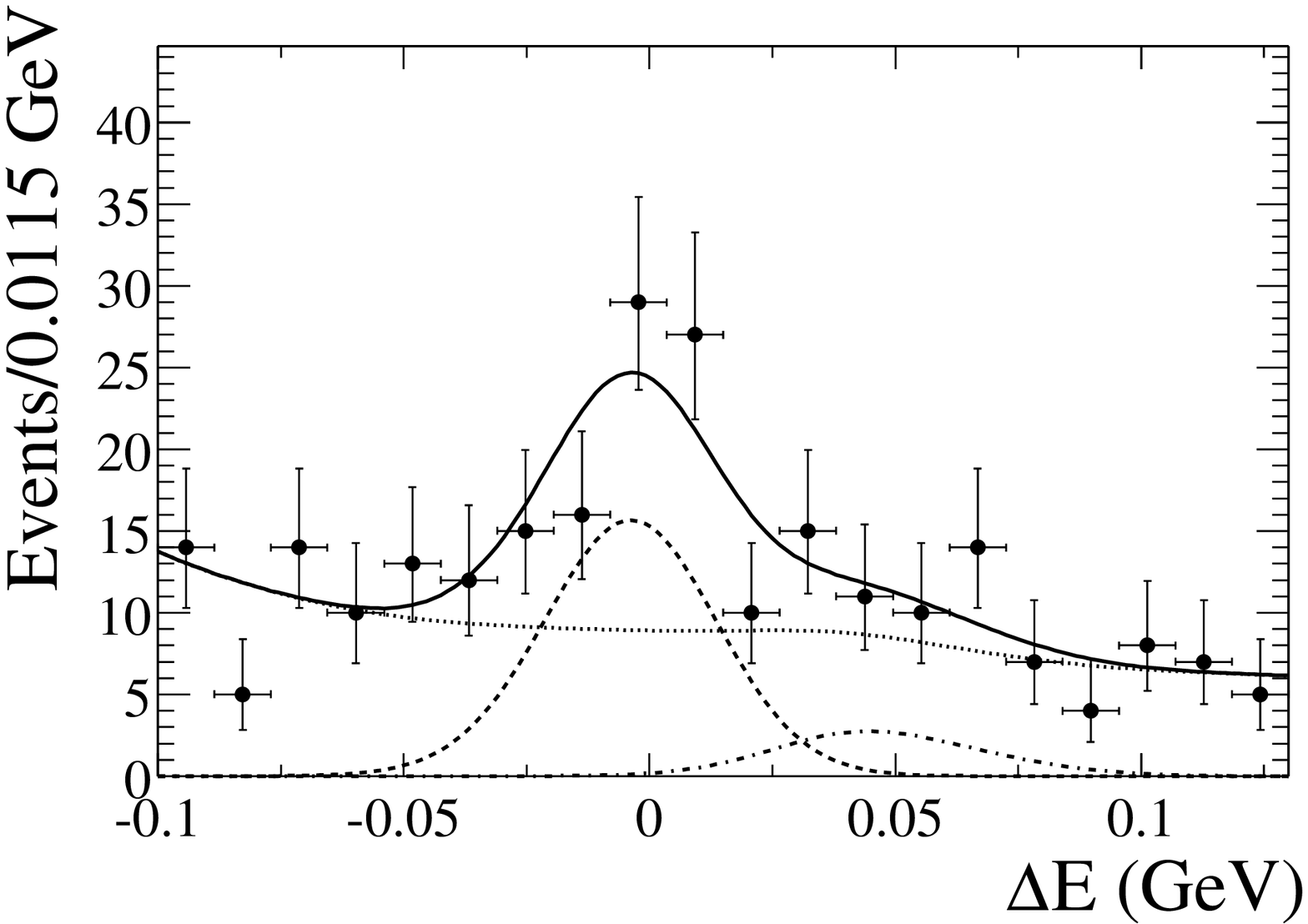}
\caption{Distributions of \DeltaE\ for events enhanced in $B\rightarrow D^0K$ signal. Top: $D^0\rightarrow K^-\pi^+, K^-\pi^+\pi^+\pi^-, K^-\pi^+\pi^0$; bottom: $D^0\rightarrow K^-K^+, \pi^-\pi^+$. Solid curves represent projections of the maximum likelihood fit; dashed, dashed-dotted and dotted curves represent the $B\rightarrow D^0K$, $B\rightarrow D^0\pi$ and background contributions.}
\label{fig:fit_noncp}
\end{center}
\end{figure}

The results of the fit are summarized in Table~\ref{tab:fitresults}.
Figure~\ref{fig:fit_noncp} shows the distributions of \DeltaE\ for the 
combined non-\CP\ and \CP\ modes after enhancing the $B\rightarrow
D^0K$ purity by requiring that the prompt track be consistent with the
kaon hypothesis and that $|\mes-\langle \mes \rangle|<3\sigma$
($|M(\pi^-\pi^+)-\langle M(\pi^-\pi^+) \rangle|<3\sigma$ for
\dotopp). 
The projection of a likelihood fit, modified 
to take into account the tighter selection criteria, is overlaid in the Figure.
 
\begin{table}[h]
\caption{Results from the maximum-likelihood fit. For the \dotokk\ and \dotopp\ modes we quote the results for the fits performed on the whole sample
and on the $B^+$ and $B^-$ subsamples.
}
\label{tab:fitresults}
\begin{center}
\begin{tabular}{lccc}
\hline
\hline
$D^0$ mode &\  $N(B\rightarrow D^0\pi)$ &\ $N(B\rightarrow D^0K)$ &\ \ \ $N(\FourS)$\\
\hline
$K^-\pi^+$           &\  $4440\pm 69$ &\  $360 \pm 21$ &\ \ \ $61.0\times 10^{6}$\\
$K^-\pi^+\pi^+\pi^-$ &\  $2914\pm 56$ &\  $242 \pm 18$ &\ \ \ $61.0\times 10^{6}$\\
$K^-\pi^+\pi^0$      &\  $2650\pm 56$ &\  $208 \pm 18$ &\ \ \ $61.0\times 10^{6}$\\ 
\hline
$K^-K^+$         &\  $565 \pm 25$ &\  $44.3 \pm 9.0$ &\ \ \ $88.8\times 10^{6}$\\
$K^-K^+$ [$B^+$] &\  $286 \pm 18$ &\  $16.7 \pm 5.8$ &\ \ \ $88.8\times 10^{6}$\\
$K^-K^+$ [$B^-$] &\  $280 \pm 18$ &\  $27.8 \pm 6.8$ &\ \ \ $88.8\times 10^{6}$\\
\hline
$\pi^-\pi^+$         &\  $195 \pm 17$ &\  $24.2 \pm 7.2$       &\ \ \ $88.8\times 10^{6}$\\
$\pi^-\pi^+$ [$B^+$] &\  $99 \pm 12$  &\  $16.8^{+5.6}_{-4.9}$ &\ \ \ $88.8\times 10^{6}$\\
$\pi^-\pi^+$ [$B^-$] &\  $96 \pm 12$  &\  $6.5^{+5.1}_{-4.3}$  &\ \ \ $88.8\times 10^{6}$\\
\hline
\hline
\end{tabular}
\end{center}

\end{table}

The ratios $R$ and $R_{\CP}$ are computed by scaling the ratios of the numbers 
of \btodk\ and \btodp\ mesons 
by correction factors that account for small differences in the efficiency 
between \btodk\ and \btodp\ selection, estimated with simulated signal 
samples. The results are listed in Table~\ref{tab:final_ratio}. 

The direct \CP\ asymmetry $A_{\CP}$
for the \btodkcp\ decays is calculated from the measured
yields of positive and negative charged meson decays reported in
Table~\ref{tab:fitresults}. 
We measure $A_{\CP}= 0.07\pm 0.17\stat\pm 0.06\syst$.

Systematic uncertainties in the ratios $R$, $R_{\CP}$ and in the \CP\ 
asymmetry $A_{\CP}$ arise primarily from uncertainties in signal yields 
due to imperfect knowledge of the PDF shapes. 
The parameters of the analytical PDFs are varied by $\pm1\sigma$ 
and the difference in the signal yields 
is taken as a systematic uncertainty. 
When a \BB\ PDF is parameterized through the kernel estimation, we repeat the 
fit using several statistically independent simulated \BB\ samples to define 
the PDF. The width of the distribution of the difference between 
the new yields and the original yield is taken as the systematic uncertainty.

The uncertainties in the branching fractions of the channels 
contributing to the \BB\ background have been taken into account.
The correlations between the different sources of systematic errors, when
non-negligible, are considered. An upper limit on intrinsic detector charge
bias due to acceptance, tracking, and particle identification
efficiency 
has been obtained from the measured
asymmetries in the processes $B^-{\ra}D^0[{\ra}K^-\pi^+]h^-$ and 
$B^-{\ra}D^0_{\CP}\pi^-$, where \CP\ violation is expected to be negligible.
This limit (0.04) has been added in quadrature to the total systematic
uncertainty on the \CP\ asymmetry. 

\begin{table}[h]
\caption{Measured ratios $R$ and $R_{\CP}$
for different \Dz\ decay modes. The first error is statistical, the second is
systematic.} 
\label{tab:final_ratio}
\begin{center}
\begin{tabular}{lc}
\hline
\hline
\btodh\ decay mode &\ \ \ \ \ \ \textbf{\BR($B{\ra}DK$)/\BR($B{\ra}D\pi$)} $(\%)$\\
\hline
\hline
\dotokp&\ \ \ \ \ \  $8.4\pm 0.5 \pm 0.2$\\
\dotokppp&\ \ \ \ \ \  $8.7\pm 0.7 \pm 0.2$\\
\dotokppo&\ \ \ \ \ \  $7.7\pm 0.7 \pm 0.3$\\
\hline
weighted mean&\ \ \ \ \ \  $8.31\pm 0.35\pm 0.20$\\
\hline
\hline
\dotokk &\ \ \ \ \ \  $8.0\pm 1.7 \pm 0.6$\\
\dotopp &\ \ \ \ \ \  $12.9\pm 4.0^{+1.1}_{-1.5}$\\
\hline
weighted mean&\ \ \ \ \ \  $8.8\pm 1.6 \pm 0.5$\\
\hline
\hline
\end{tabular}
\end{center}
\end{table}

In conclusion, we have reconstructed \btodk\ decays with
$D^0$ mesons decaying to non-\CP\ and \CP-even eigenstates.
The ratios $R_{(\CP)}$ of the branching fractions
$\BR(B^-{\ra}D^0_{(\CP)} K^-)$ and $\BR(B^-{\ra}D^0_{(\CP)} \pi^-)$
and the direct \CP\ asymmetry $A_{\CP}$ have been measured.
The measured ratio $R$ is consistent with Standard Model
expectation ($\approx 7.5\%$) assuming factorization~\cite{bib:RinMS}. 
In the Standard Model $R_{\CP}/R=1+r^2+2r\cos\delta\cos\gamma$ and
$A_{\CP}=2r\sin\delta\sin\gamma/(1+r^2+2r\cos\delta\cos\gamma)$, where  
$r\approx0.1-0.2$ is the magnitude of the ratio of the amplitudes for
the processes $B^-\ra \Dzb K^-$ and $B^-\ra D^0 K^-$, and
$\delta$ is the (unknown) relative strong phase between these two
amplitudes~\cite{gronau1991}.  
The measured values of $R$ and $R_{\CP}$ are equal
within errors, and $A_{\CP}$ is consistent with zero. These
results, together with the ones obtained by CLEO and 
Belle~\cite{bellebtdk}, represent a first step towards the
measurement of the angle $\gamma$ and of direct {\CP} violation in the
$B$ system using the \btodk\ decays. 

We are grateful for the excellent luminosity and machine conditions
provided by our \pep2\ colleagues, 
and for the substantial dedicated effort from
the computing organizations that support \babar.
The collaborating institutions wish to thank 
SLAC for its support and kind hospitality. 
This work is supported by
DOE
and NSF (USA),
NSERC (Canada),
IHEP (China),
CEA and
CNRS-IN2P3
(France),
BMBF and DFG
(Germany),
INFN (Italy),
FOM (the Netherlands),
NFR (Norway),
MIST (Russia), and
PPARC (United Kingdom). 
Individuals have received support from the 
A.~P.~Sloan Foundation, 
Research Corporation,
and Alexander von Humboldt Foundation.

%
%

\end{document}